\documentclass[3p,times,procedia]{elsarticle}
\usepackage{nupha_ecrc}
\usepackage{units}
\usepackage[utf8]{inputenc}

\volume{00}

\firstpage{1}

\journalname{Nuclear Physics A}

\runauth{J.F. Grosse-Oetringhaus}

\jid{nupha}

\jnltitlelogo{Nuclear Physics A}

\usepackage{amssymb}
\biboptions{sort&compress}

\usepackage[figuresright]{rotating}

\newcommand{\RCP}          {\ensuremath{R_{\rm CP}}}
\newcommand{\RAA}          {\ensuremath{R_{\rm AA}}}

\newcommand{\Npart}        {\ensuremath{N_{\rm part}}}

\newcommand{\dNdeta}       {\mathrm{d}N_\mathrm{ch}/\mathrm{d}\eta}

\newcommand{\pt}           {\ensuremath{p_{\mathrm{T}}}}

\newcommand{\snn}          {\ensuremath{\sqrt{s_{\mathrm{NN}}}}}

\newcommand{\com}[1]       {}

\begin{document}

\begin{frontmatter}



%% Instructions from Editor: Please use the following \dochead only in the preprint version (e-print arXiv etc.);
%% use empty \dochead{} when submitting to Nuclear Physics A!
\dochead{XXVIIth International Conference on Ultrarelativistic Nucleus-Nucleus Collisions\\ (Quark Matter 2018)}
%% e.g. \dochead{17th International Conference on Dynamical Processes in Excited States of Solids}

\title{The Future of High-Energy Heavy-Ion Facilities}


\author{Jan Fiete Grosse-Oetringhaus}

\address{CERN, 1211 Geneva 23, Switzerland}

\begin{abstract}
The plans within the next decade of the high-energy heavy-ion facilities RHIC at BNL and LHC at CERN are reviewed in detail, focusing on the physics programme for $\snn \ge \unit[200]{GeV}$. The expected data samples are presented, together with a discussion of the physics programme and reach. Selected performance studies are presented. An outlook is given on the plans with these and new facilities beyond 2030.
\end{abstract}

\end{frontmatter}

% \linenumbers

\section{Introduction}
Initially the field of heavy-ion physics searched for a new state of matter, the Quark--Gluon Plasma (QGP). Following its discovery, a precise characterization of the observed hot and dense medium followed which is still ongoing today.
Nowadays, the main open questions are on the one hand the question of the underlying dynamics of the medium, its degrees of freedom and the microscopic structure. Can one find a model which describes at the same time the long wavelength (ideal fluid) behaviour and the short wavelength (quenching) behaviour?
On the other hand, the discovery of effects usually associated with a hot and dense medium in p--Pb, d--Au, $^3$He--Au, and smaller systems down to pp, asks the question of the onset of the QGP as a function of system size. The observations challenge two paradigms at once: a) down to which system size does the currently successful theoretical description of heavy-ion collisions remain valid; and b) can the standard tools for minimum-bias pp physics remain standard? Finding a solution to these questions may be the key to finding a universal description of the underlying physics of non-perturbative QCD.

The next decade of heavy-ion physics at high energy will use rare probes to understand the inner workings and the onset of the QGP. This aim is achieved by upgrading the accelerators and in particular the detectors to study one to two orders of magnitude more collisions than currently achievable. The programme using rare probes builds on several distinct pillars:
\begin{enumerate}
\item the study of charm and beauty, in particular at low $\pt$, which are produced in the initial hard scattering and travel through the entire medium like Brownian motion markers;
\item the study of low-mass dielectrons as a window to signs of chiral-symmetry restoration and thermal radiation;
\item the study of jets and jet substructure to understand energy-loss mechanisms and scattering centres in the QGP;
\item the study of small systems to unravel the origin of collective phenomena in these systems and the onset of the QGP.
\end{enumerate}
In addition, various other topics can be studied with the unprecedented expected data samples: for example rare nuclei, top quarks as a novel way to put constraints on nuclear PDFs, and ultraperipheral collisions including light-by-light scattering.

\section{Planned Detector Upgrades at LHC and RHIC}
Both currently running heavy-ion facilities undergo significant upgrades in the next 10 years.

The \textbf{STAR} detector at RHIC currently upgrades their TPC to $|\eta| < 1.5$ and adds an additional forward TOF, as well as event plane detectors on both sides of the experiment~\cite{star_mid}. In addition, forward upgrades are considered for 2022, adding tracking and calorimeters in $-4.6 < \eta < -2.5$~\cite{star_fwd}. STAR's focus is on exploiting particle identification, polarised proton beams and forward physics.

The \textbf{sPHENIX} detector will replace PHENIX at RHIC~\cite{Adare:2015kwa}. This compact detector with silicon and TPC tracking as well as electromagnetic and hadronic calorimeters covers $|\eta| < 1.1$. From 2023, it will operate with continuous data-taking recording \unit[15]{kHz} Au--Au collisions focussing on jets, $\Upsilon$ and heavy flavour.

The \textbf{ALICE} detector at LHC undergoes a significant upgrade in the next years~\cite{Abelev:1475243, Abelev:1625842}, replacing the inner tracking system and the readout of the TPC. In addition, a silicon vertex detector for the forward muon arm and fast trigger detectors are added. Continuous data-taking recording \unit[50]{kHz} Pb--Pb collisions is planned from 2021. The focus is on untriggerable signals with tiny signal over background, in particular heavy flavour and low-mass dileptons.

The \textbf{CMS} detector at LHC plans to upgrade their inner tracker in 2024/25 enlarging the acceptance to $|\eta| < 4$~\cite{Contardo:2020886}. Similarly and on the same time scale, the \textbf{ATLAS} detector at LHC will upgrade their inner tracker to an acceptance of $|\eta| < 4$ and add a timing detector for pile up detection~\cite{CERN-LHCC-2015-020}. In addition, both experiments will improve the granularity of the forward calorimetry as well as with the trigger and DAQ. Their focus in heavy-ion collisions will be on heavy flavour, jets and $\Upsilon$.

In 2020/21, the \textbf{LHCb} detector at LHC will replace their tracker to achieve 10 times better granularity required for reconstructing heavy-ion collisions~\cite{Collaboration:1647400}. Event selection will be fully software based, and a storage cell is considered to be added for an enhanced fixed-target programme. Under discussion are further upgrades in mid-2020 (TOF) and end-2020 (further occupancy increase)~\cite{Bediaga:2018lhg}. The potential in heavy-ion collisions needs to be evaluated in detail but inspecting Pb--Pb collisions at a rate of \unit[20]{kHz} seems feasible.

\begin{figure}[b!]
\centering
\includegraphics[width=0.9\linewidth]{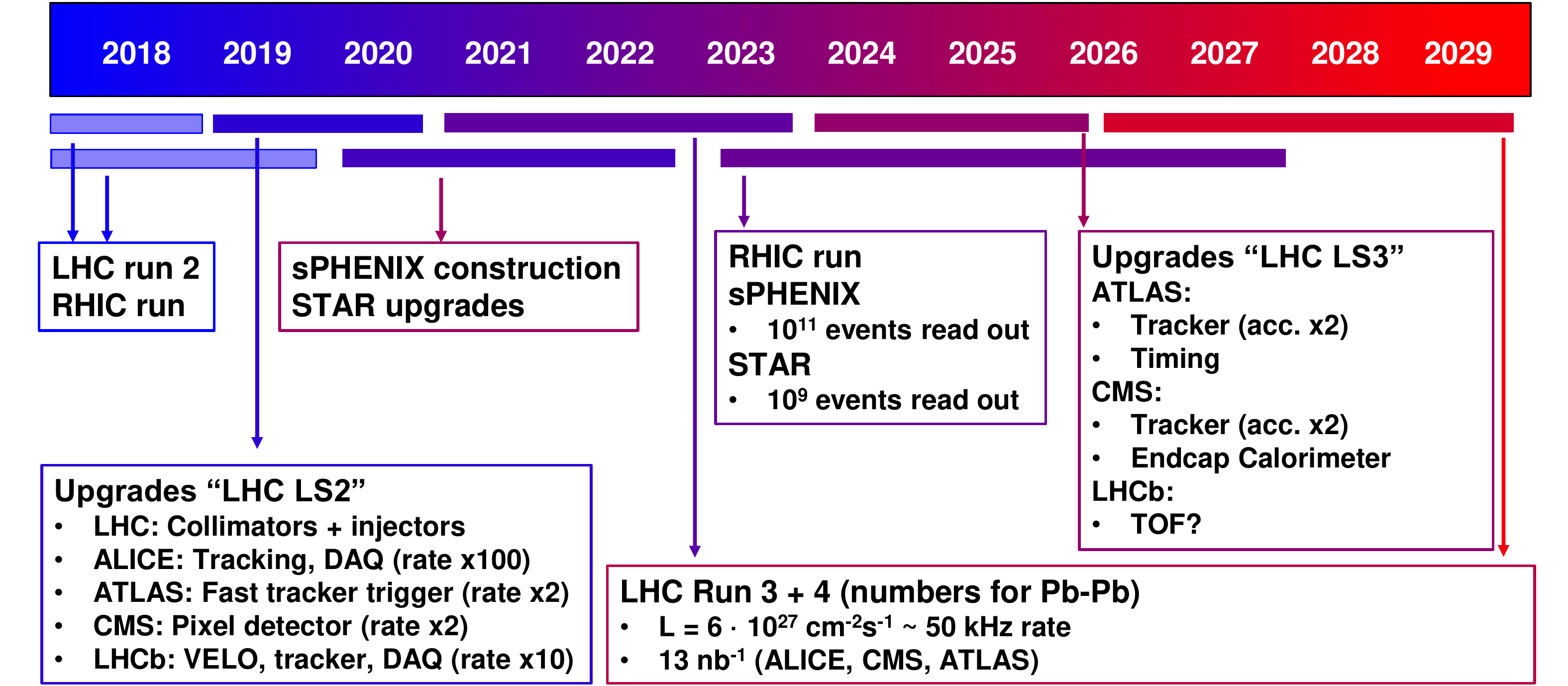}
\caption{Timeline of the planned detector upgrades and data-taking periods.}
\label{fig:timeline}
\end{figure}

Figure~\ref{fig:timeline} presents a timeline of the planned detector upgrades and data-taking periods. The current LHC (RHIC) runs will end in 2018 (2019), and upgrade periods will follow. Run 3 and 4 at the LHC will take place from 2021 to 2023 and 2026 to 2029, respectively. RHIC plans to restart data-taking from 2023.

\section{Expected Performance}

The planned upgrades and facilities increase the available data by up to two orders of magnitude. At LHC, \unit[13]{nb$^{-1}$} of Pb--Pb collisions are planned to be delivered. These will be fully recorded by the ALICE detector leading to an unprecedented sample of $10^{11}$ minimum-bias Pb--Pb collisions. The same luminosity will be fully sampled for hard probes by the triggers of CMS and ATLAS. sPHENIX will record about $10^{11}$ minimum-bias Au--Au collisions.

These numbers will allow precision measurements of the rare and hard probes sector of heavy-ion collisions ranging from $D$ and $B$ mesons, $\Upsilon$ states as well as jets, to the measurement of $\Lambda_c$, $\Lambda_b$ and $^4_\Lambda H$. The remainder of this manuscript gives some highlights of the expected performance, and can obviously only scratch the surface of this rich programme. For more details, the reader is referred to Refs.~\cite{star_mid, star_fwd, Adare:2015kwa, Abelev:1475243, Abelev:1625842, CMS-PAS-FTR-13-025, ATLAS-Collaboration:1472528, Bediaga:2018lhg}.

\subsection{Heavy Flavour and Quarkonia}

\begin{figure}[t!]
\centering
\includegraphics[width=0.49\linewidth]{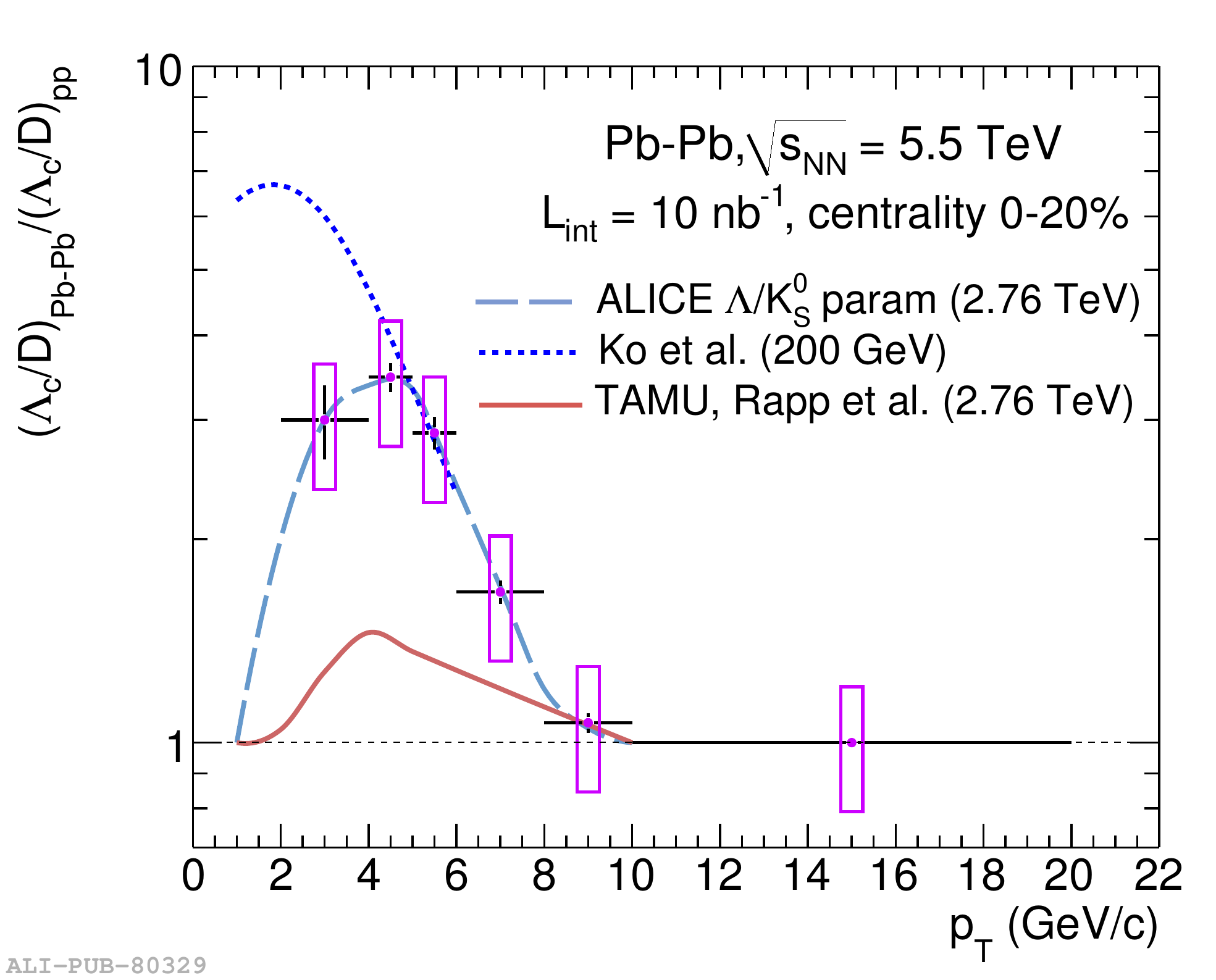}
\caption{Expected performance of the $\Lambda_c$ over $D$ ratio as a function of $\pt$ measurement by ALICE, compared to model expectations. Figure from Ref.~\cite{Abelev:1625842}.}
\label{fig:lambda_c}
\end{figure}

\begin{figure}[t!]
\centering
\includegraphics[width=0.45\linewidth]{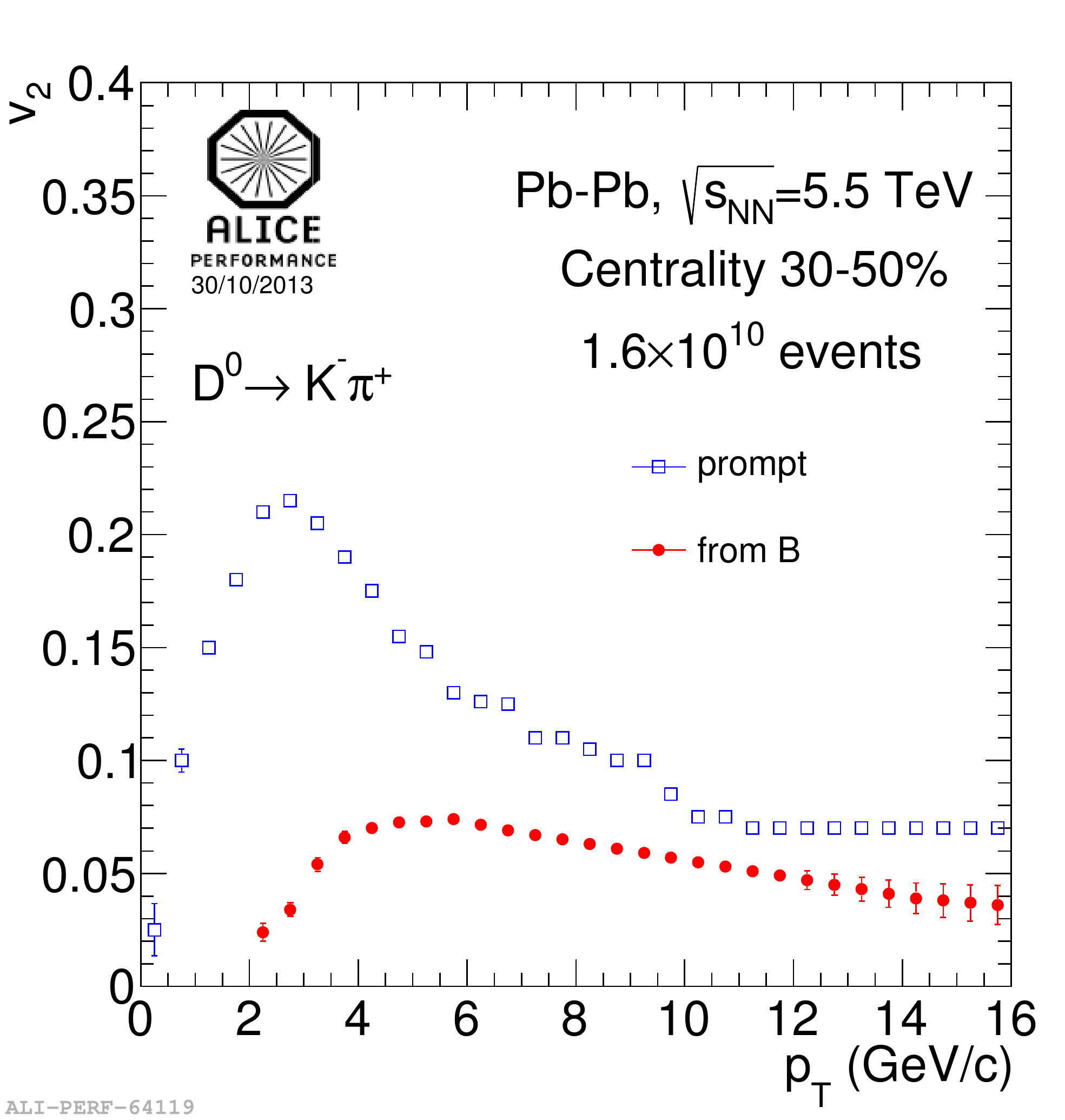}
\hfill
\includegraphics[width=0.48\linewidth]{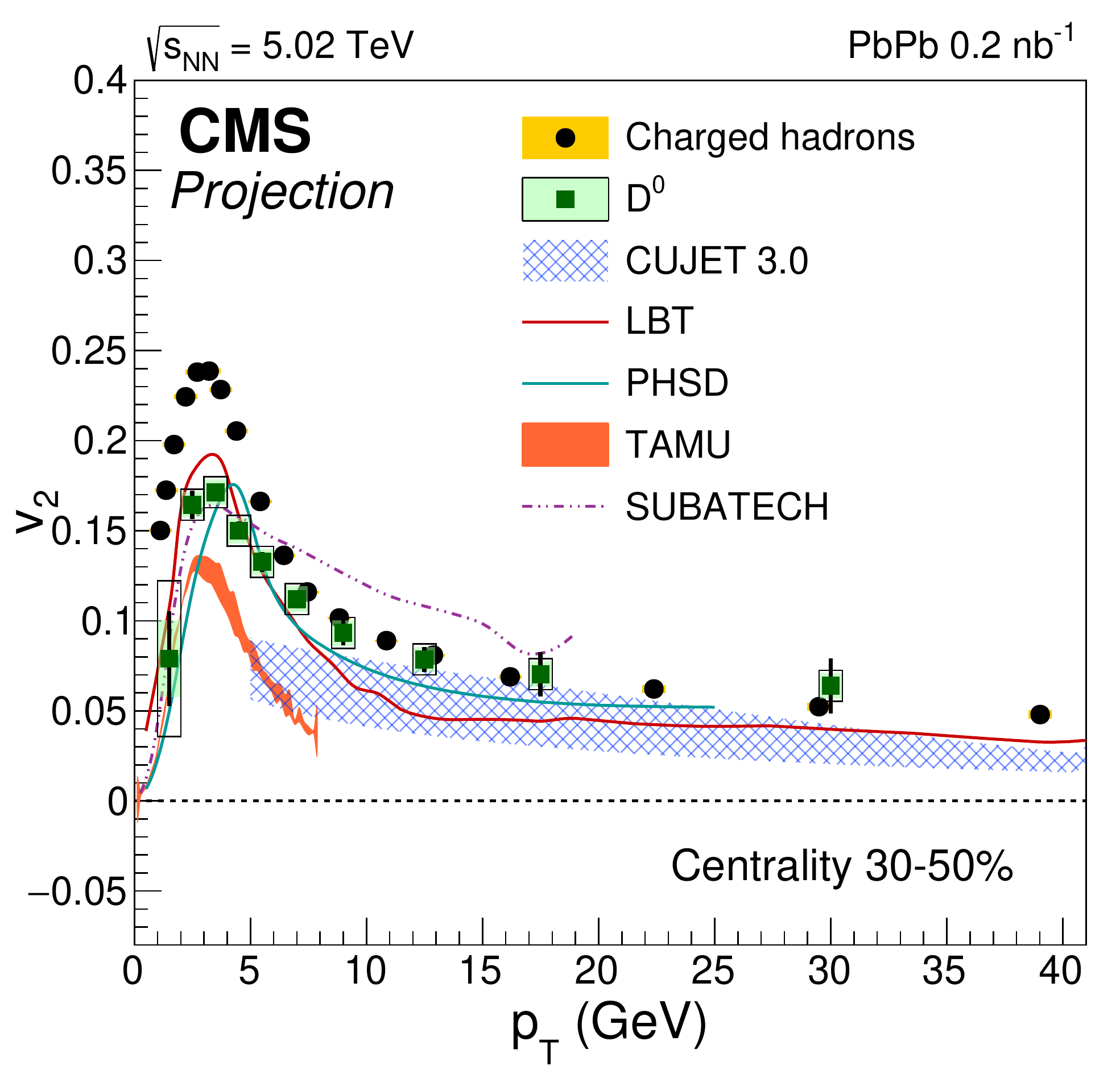}
\caption{Expected performance of the $v_2$ as a function of $\pt$ for prompt and non-prompt $D$ mesons by ALICE (left panel) and CMS (right panel). Figures from Refs.~\cite{Abelev:1625842, CMS-PAS-FTR-13-025}.}
\label{fig:v2}
\end{figure}

A key element of the future physics programme is the detailed understanding of the production of heavy-flavour baryons and mesons, the interaction of charm and beauty with the medium, and the measurement of the total charm cross-section.
Figure~\ref{fig:lambda_c} presents the expected performance of the $\Lambda_c$ over $D$ ratio compared between Pb--Pb and pp collisions. The baryon over meson ratio gives insight into how charmed hadrons recombine with light quarks at hadronization and is sensitive to radial flow. Similarly, for instance, the $D_s/D$, and $\Lambda_b/B$ ratio can be measured.
The measurement of the $D$ elliptic flow $v_2$ addresses the question of heavy quark-medium interactions and the degree of thermalization in the medium. Experimentally, the measurement of prompt and non-prompt $D$ mesons allows to measure both, the prompt $D$ production as well as the one stemming from $B$ decays. Figure~\ref{fig:v2} presents the expected performance. The measurement will shed light on the degree of correlation of light and heavy quark flow, and constrain the temperature dependence of the charm diffusion coefficient, a key QGP property.

Abundant quarkonia states will allow precision measurements of charmonia and bottomonia $R_{\rm AA}$ and $v_2$. Constraints are expected on regeneration models whose uncertainties will in addition be reduced by the measurement of the total charm cross-section, and the precision of e.g. the $J/\Psi$ $v_2$ will reach percent level.
Up to 320\,000 $\Upsilon$ are expected within the CMS acceptance; the expected performance for the three different $\Upsilon$ states can be seen in Fig.~\ref{fig:upsilon} (left). 3S suppression will be observed for the first time and, potentially, the difference between 2S and 3S suppression can be measured.

\subsection{Jets}

\begin{figure}[t!]
\centering
\includegraphics[width=0.50\linewidth]{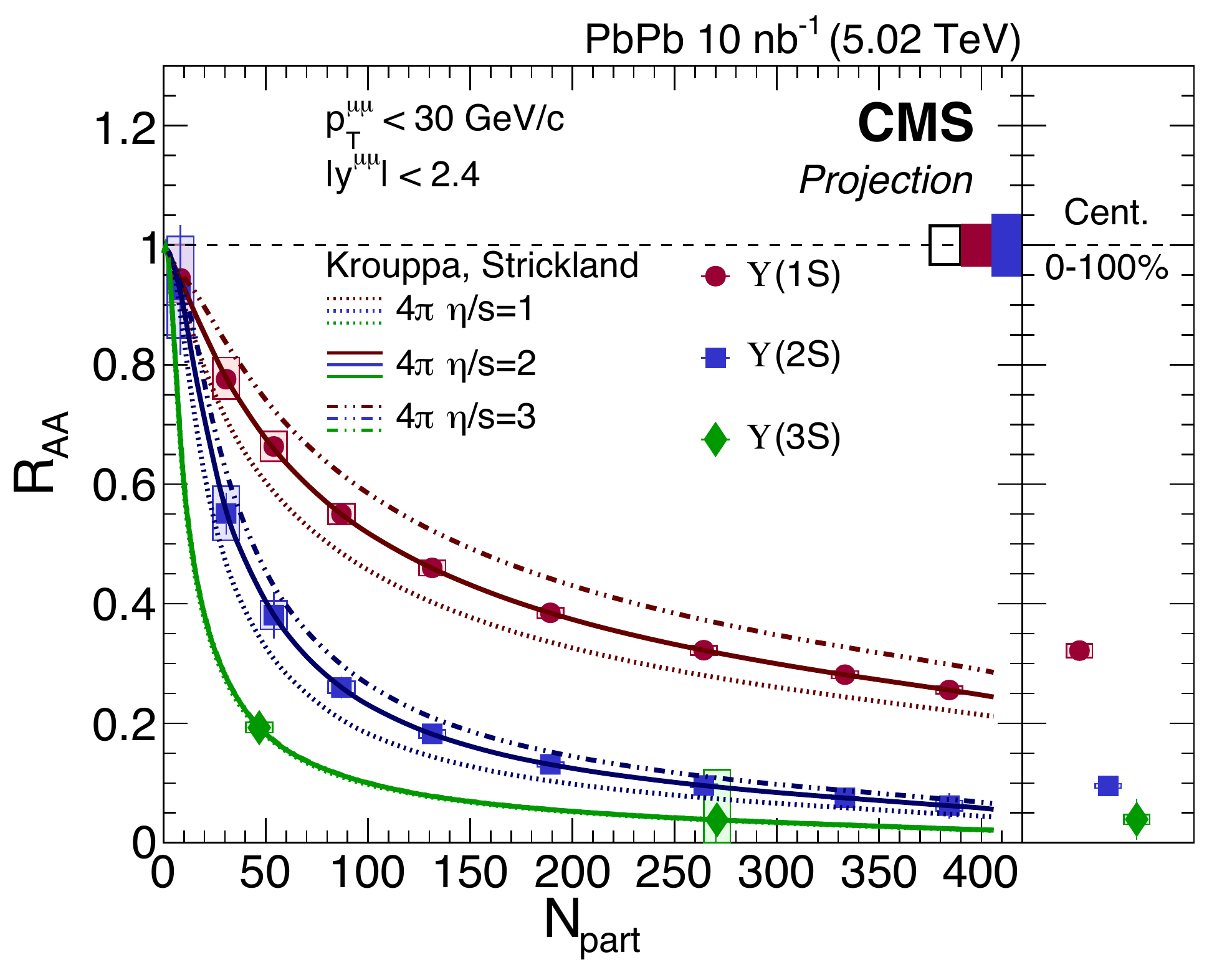}
\hfill
\includegraphics[width=0.45\linewidth]{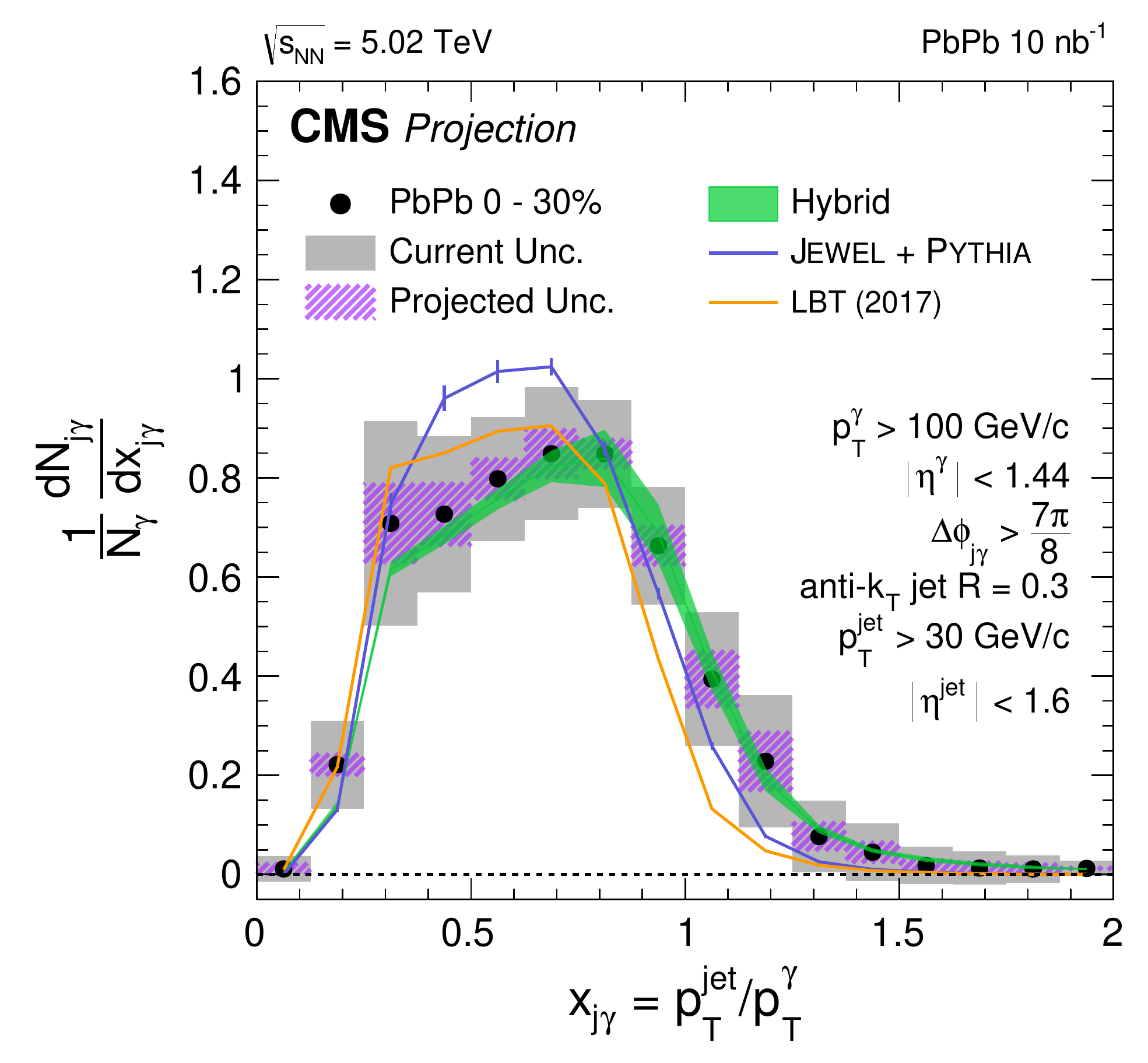}
\caption{Left panel: Expected performance of the $\Upsilon$(1S), 2S and 3S $\RAA$ as a function of $\Npart$ by CMS. Right panel: Projection of the CMS measurement of $x_{j\gamma}$ compared to the current precision, and three different model calculations. Figures from Ref.~\cite{CMS-PAS-FTR-13-025}.}
\label{fig:upsilon}
\end{figure}

The study of jets will allow precision medium tomography in the next decade. In the light-quark sector hadron and jet suppression can be measured over three orders of magnitude, into the TeV regime. The path-length dependence of energy loss can be assessed with event-shape engineering. Precision measurements are expected for example for di-jet imbalance, $\gamma$-jet and $Z$-jet correlations. Figure~\ref{fig:upsilon} (right) shows the expected precision for the $\gamma$-jet yield as a function of momentum imbalance $x_{j\gamma} = p_{\rm T}^{jet}/p_{\rm T}^{\gamma}$, directly sensitive to the lost energy in the medium.

\begin{figure}[t!]
\centering
\hfill
\includegraphics[width=0.46\linewidth]{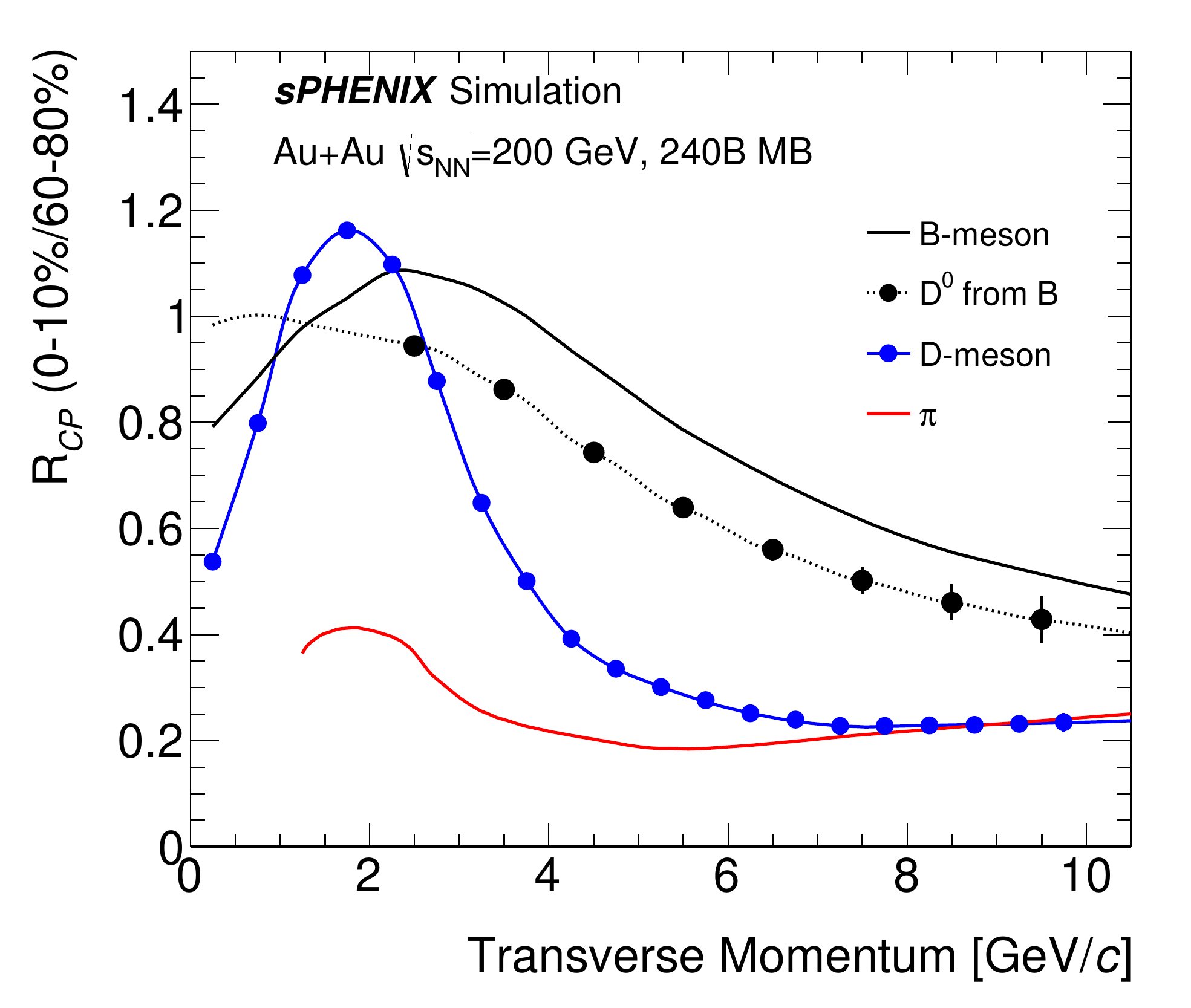}
\hfill
\includegraphics[width=0.46\linewidth]{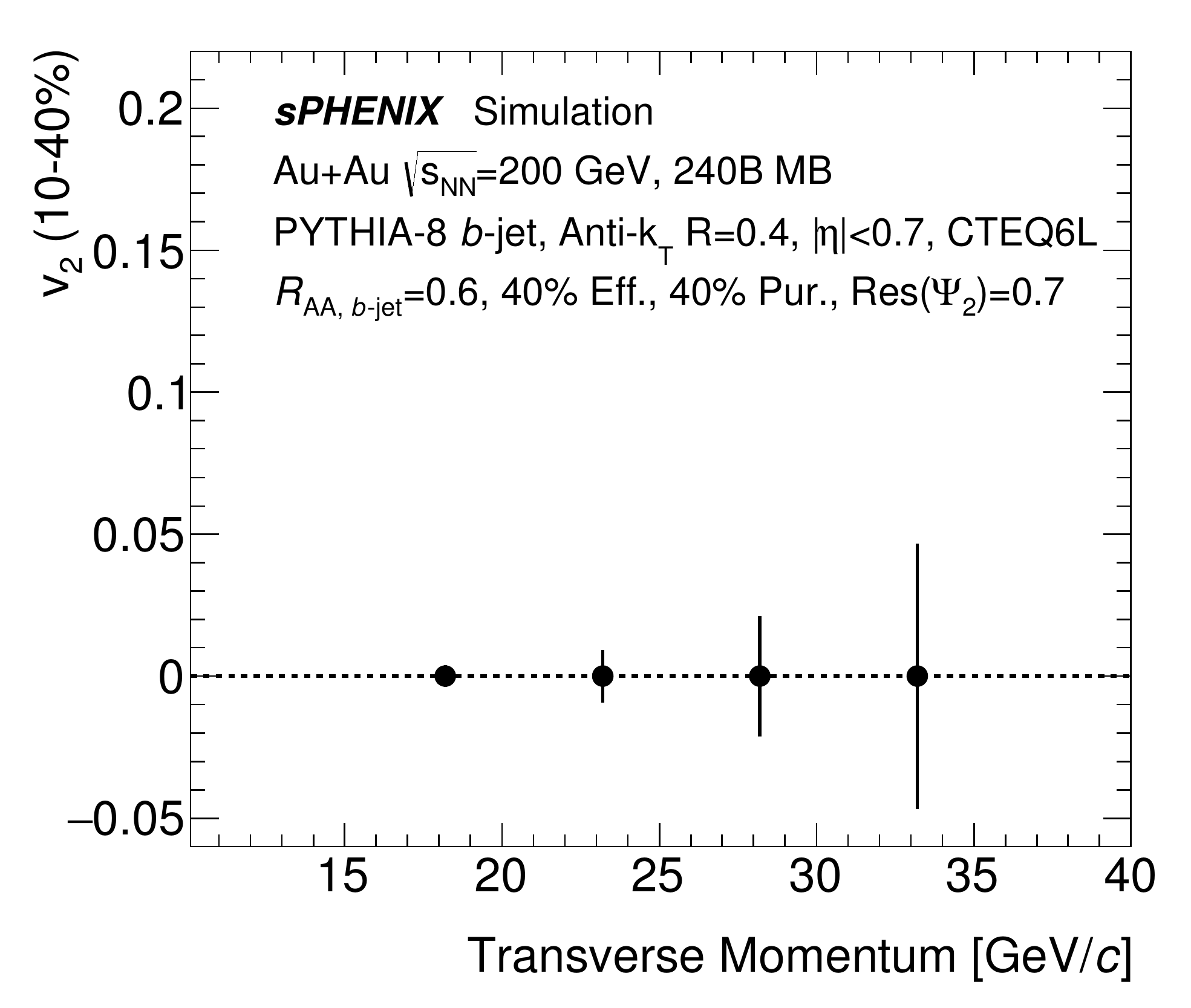}
\hfill
\caption{Left panel: Expected performance of the sPHENIX $\RCP$ of prompt and non-prompt $D$ mesons as a function of $\pt$. Right panel: Projection of the sPHENIX measurement of the $v_2$ of $b$-tagged jets as a function of $p_T$. Figures from Ref.~\cite{sphenix_mvtx}.}
\label{fig:jets}
\end{figure}

\begin{figure}[t!]
\centering
\hfill
\includegraphics[width=0.48\linewidth]{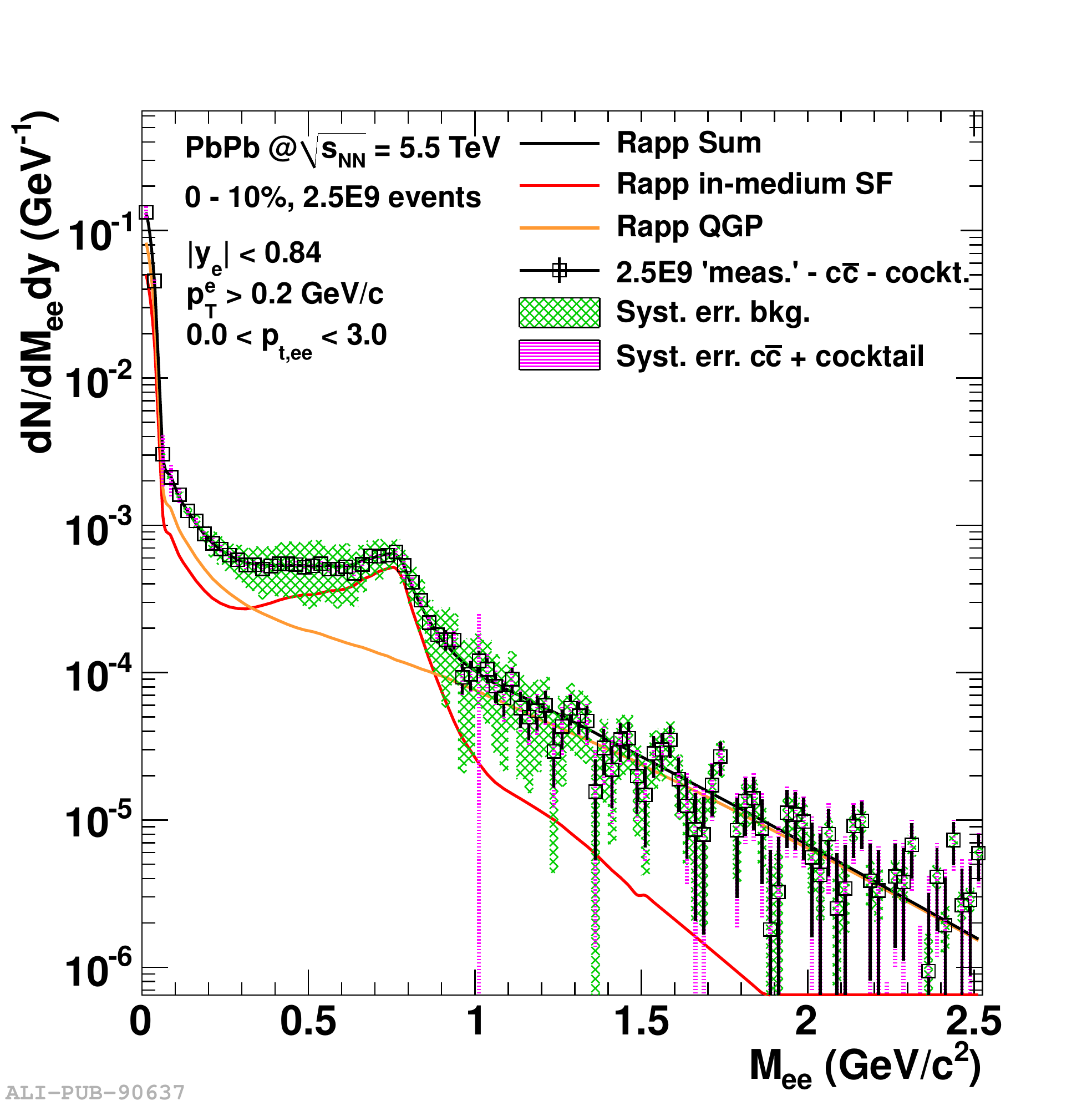}
\hfill
\includegraphics[width=0.40\linewidth]{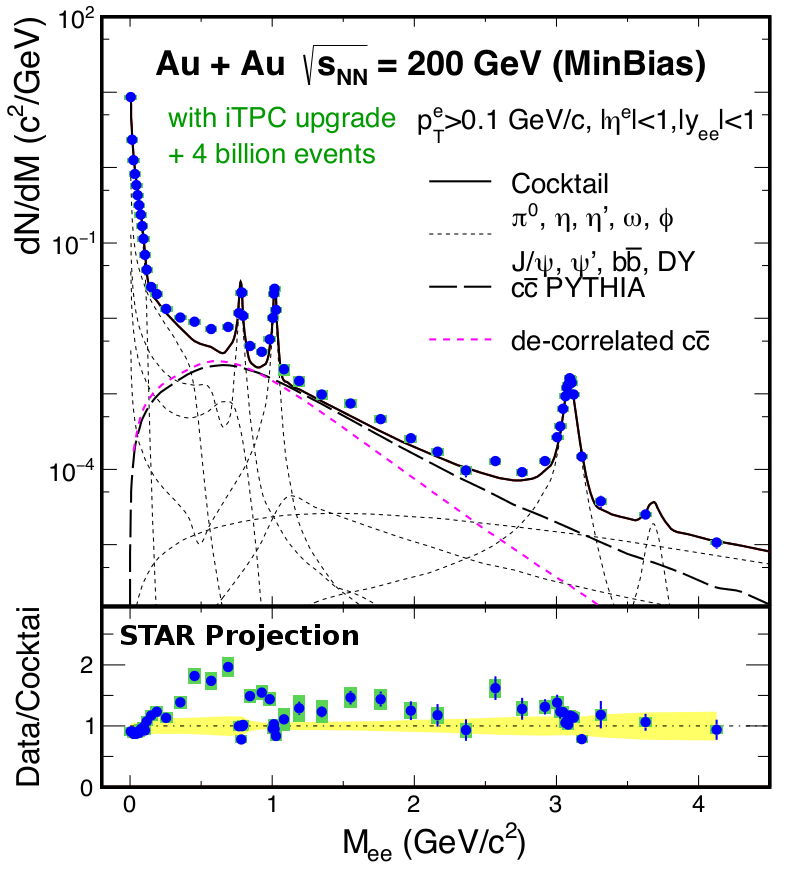}
\hfill
\caption{Expected performance of the di-electron invariant mass distribution by ALICE for 0--10\% central Pb--Pb collisions~\cite{Abelev:1625842} (left panel) and STAR for minimum-bias Au--Au collisions~\cite{star_mid} (right panel). As the statistical uncertainties are driven by the combinatorial background, the uncertainties are smaller for the STAR case than for the ALICE case. This background increases with $\sqrt{s_{\rm NN}}$; furthermore minimum-bias collisions are compared to central events, where the combinatorial background is larger but also the medium effects stronger.}
\label{fig:lmee}
\end{figure}

In the heavy-quark sector, the $R_{\rm AA}$ of $D$ and $B$ allows a precise assessment of the flavour dependence of energy loss. The expected performance is shown in Fig.~\ref{fig:jets} (left). In addition, the measurement down to $p_T = 0$ allows to study the transport of $b$ quarks in the medium. Heavy-flavour identified fragmentation functions are in reach. The path-length dependence of energy loss can be studied by measuring jet $v_2$, illustrated for $b$-tagged jets in Figure~\ref{fig:jets} (right).

Jet substructure observables, well established in the high-energy physics community, promise a differential probing of the parton-splitting phase space. With the associated methods, like grooming, regions can be isolated where medium affects are strongest. An interesting direction is also to study large-angle Moli{\`e}re scattering as a window to identify point-like quark and gluon scattering centers at large $Q^2$.

\subsection{Low-mass dileptons}

The potential of the measurement of low-mass dileptons is unprecedented in the next decade, in particular due to the low-material budget of the upgraded ALICE detector and special running with low magnetic field. Figure~\ref{fig:lmee} presents the expected performance of the di-electron invariant-mass distribution for ALICE and STAR.
This distributions allows to measure the $\rho$ meson spectral shape which is expected to be modified due to chiral symmetry restoration in the Quark--Gluon Plasma which in turn drives the generation of hadron masses. Further, the invariant-mass distribution encodes the temperature of the emitting medium, as thermal dileptons are not blueshifted, as well as giving access to the space-time evolution of the system. Expected uncertainties at LHC are 10--25\% on the temperature extraction, 1\% absolute on the elliptic flow $v_2$ and 1--2\% on $R_\gamma$, the ratio of the inclusive photon yield to the decay photon yield.

\subsection{Small Systems}

\begin{figure}[t]
\centering
\includegraphics[width=0.46\linewidth]{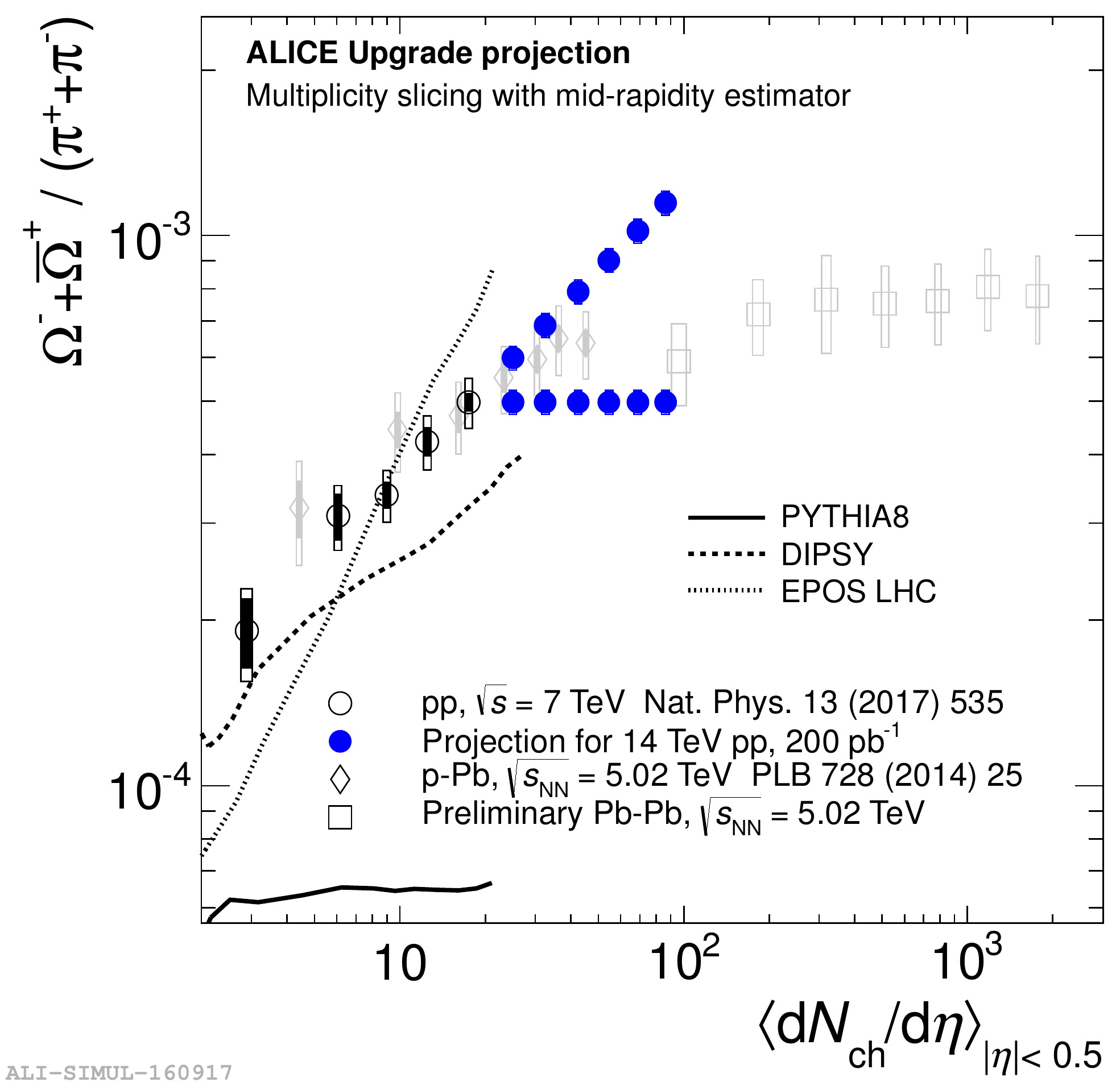}
\hfill
\includegraphics[width=0.49\linewidth]{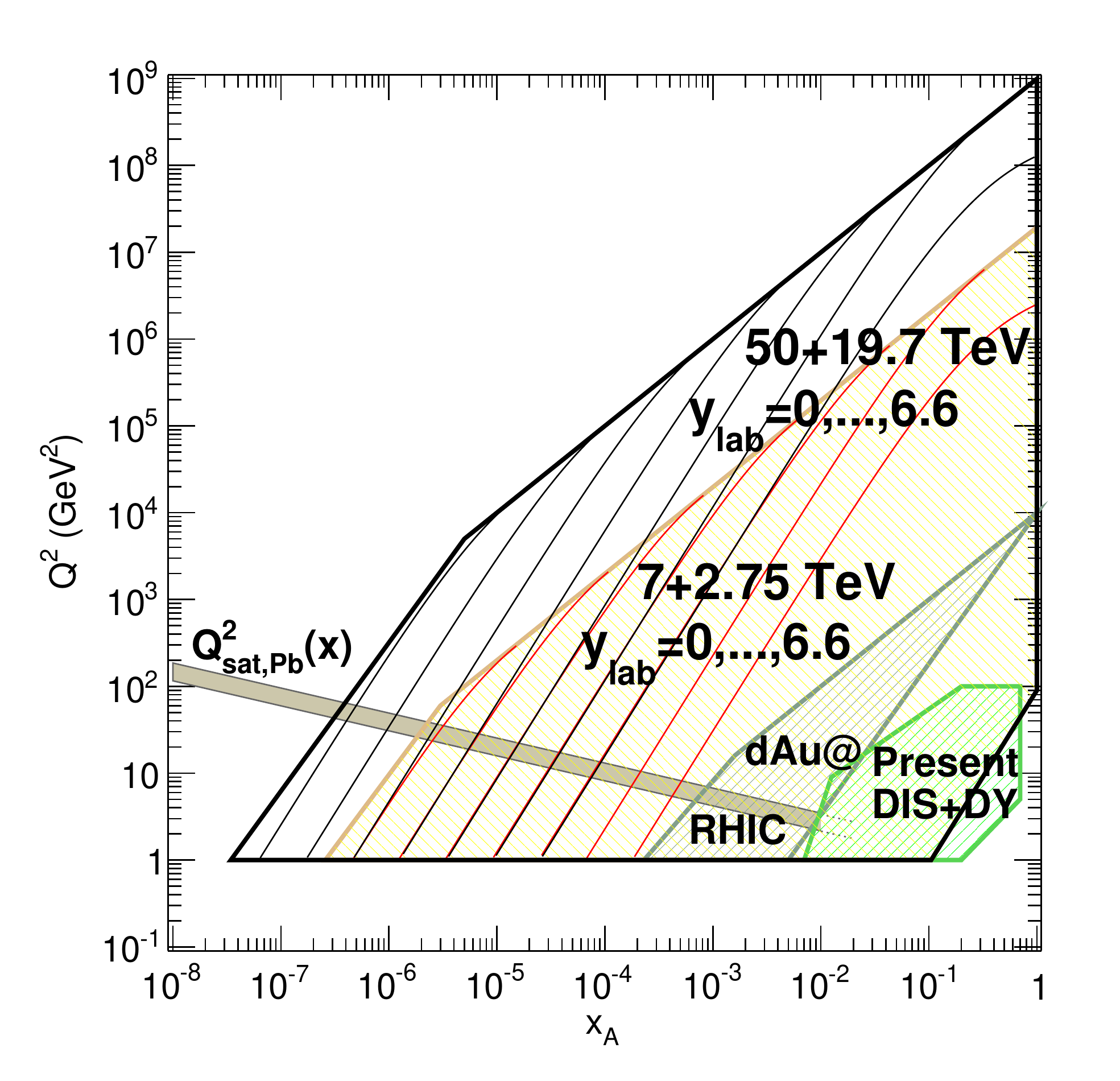}
\caption{Left panel: $\Omega/\pi$ ratio as a function of multiplicity in pp, p--Pb, and Pb--Pb collisions (data from Ref.~\cite{ALICE:2017jyt}) overlaid with the expected ALICE performance for a \unit[200]{pb$^{-1}$} high-multiplicity pp programme.
Right panel: $x$--$Q^2$ plane explorable in proton--nucleus collisions at FCC, LHC and RHIC compared to existing DIS and DY data. Figure from Ref.~\cite{Dainese:2016gch}.}
\label{fig:smallsystems}
\end{figure}

The observation of collective effects in small collision systems like p--Pb, d--Au, $^3$He--Au, and pp collisions have caused a paradigm shift. At present the underlying physics of the observed collective effects is not understood, and important experimental questions are open. An extensive programme is planned in the next decade at the LHC sampling very high multiplicity pp events ($L_{\rm int} = \unit[200]{pb^{-1}}$). The expected multiplicities are expected to be as large as the multiplicity at about 65\% central Pb--Pb collisions resulting in a qualitative new feature: a large overlap between pp and Pb--Pb collisions.
%Furthermore, further p--Pb running will significantly enhance the reach in the p--Pb system as well.
Figure~\ref{fig:smallsystems} presents the expected performance of the measurement of the $\Omega/\pi$ ratio with two scenarios extrapolating the observed increase of the ratio in pp collisions. The important question can be answered if the thermal limit of production in pp is reached as it is in Pb--Pb collisions. Further, this unprecedented data sample will allow to measure the probability distribution of $v_n$ coefficients, thermal radiation, and possibly heavy-flavour meson flow in pp collisions. The question if there are sufficient final-state interactions to cause energy loss in pp collisions can be addressed with correlation measurements ($h$--jet, dijet, $\gamma$--jet, $Z$--jet).

\subsection{Small x, Saturation, and Ultraperipheral Collisions}

A large kinematic region is accessible in p--Pb collisions at the LHC and d--Au collisions at RHIC. Figure~\ref{fig:smallsystems} (right) presents the $x$--$Q^2$ plane covered compared to existing DIS and DY data. The reach extends well into the saturation regime. The current constraints on nuclear PDF fits are rather limited, and will greatly benefit from the upgraded detectors and large rates. Large-statistics measurements of dijets, photon--jet correlations, $W$, $Z$ and Drell-Yan are expected.

Ultraperipheral heavy-ion collisions offer an interesting opportunity to study nuclear modifications of the PDFs as they provide clean photon--nucleus interactions. For example, in dijet production the underlying event activity is significantly suppressed compared to p--Pb collisions allowing to access lower jet momenta and therefore smaller $Q^2$ and $x$ where the nPDF uncertainties are large.

%Ultra-peripheral heavy-ion collisions provide an opportunity to study nuclear modifications of the PDFs in clean photon-nucleus interactions. One possible observable is dijet production as suggested in Ref.~\cite{Strikman:2005yv}. Compared to the dijet production in p+Pb collisions the photo-nuclear events have less underlying event activity since multiparton interactions are significantly suppressed. This enables jet reconstruction at lower transverse momenta allowing to study nPDFs at smaller $Q^2$ and $x$ where the current uncertainties are more pronounced.

% Discuss also nuclei! CME?

%LambdaC/D      ALICE
%D v2           ALICE + CMS?
%Upsilon RAA    CMS
%b RCP          sPHENIX
%gamma-jet      CMS
%bjet v2        sPHENIX
%omega/pi       ALICE
%thermal rad.   ALICE + STAR

%NPDF?

%?? J/psi v2    ALICE
%?? splitting      CMS
%?? nuclei      ALICE

\section{Beyond 2030}

The future of the RHIC accelerator depends in particular on the development of the future electron--ion collider, see Ref.~\cite{Jin} for details.
At the LHC, running with heavy-ion beams is firmly planned until about 2030. Subsequent running is currently under discussion, including in the potential HE-LHC period where a doubling of the energy is considered with an upgraded accelerator in the LHC tunnel.

Two future large projects are under discussion aiming at a \unit[80--100]{km} circumference machine providing nucleus--nucleus collisions at about \unit[39]{TeV} and proton--nucleus collisions at \unit[63]{TeV}: the future circular collider at CERN~\cite{Dainese:2016gch} and the SPPC~\cite{Chang:2015hqa} in China.
These projects, if realized, could be available in 2040--50.
At \unit[39]{TeV}, heavy-ion collisions are expected to have a $\dNdeta$ of about 3600, a freeze-out volume of approximately \unit[11\,000]{fm$^3$}, an initial energy density at $\tau = \unit[1]{fm/\emph{c}}$ of \unit[35--40]{GeV/fm$^3$} and a life time of about \unit[13]{fm/$c$}. The FCC could collect \unit[35--110]{nb$^{-1}$} Pb--Pb collisions per month which is about 12--40 times more than LHC in future years.
At such centre of mass energies, about 30\% of the charm may be thermally produced, and the top quark becomes available as tool to study the medium. In particular, boosted tops can access its time evolution~\cite{Apolinario:2017sob}: depending on the boost, the interactions between the quarks in the decay chain $t\bar{t} \rightarrow b\bar{b} + 2W \rightarrow b\bar{b} + q\bar{q} + l + \nu$ and the medium is delayed. Thus, for the first time, the evolution of the quenching power of the medium with time can be observed. Furthermore, proton--nucleus collisions are a promising complementary programme to an electron--ion collider to study saturation phenomena at low $x$ (see Fig.~\ref{fig:smallsystems}, right) measured for example through photon production, photon--hadron correlations at forward rapidity and Drell-Yan.

\vspace{0.5cm}
The author looks forward to a rich and interesting future for heavy-ion physics.

% time evolution of the QGP -> 1711.03105

% \label{}



%% References


\bibliographystyle{elsarticle-num}
\bibliography{JFGO}






% \end{thebibliography}

\end{document}